\documentclass[
aps, prd, reprint,
10pt, notitlepage, a4paper,
floats, floatfix,
amsmath, amssymb, amsfonts, eqsecnum,
superscriptaddress,
showpacs, showkeys,
nofootinbib,
longbibliography,
]{revtex4-2}

\usepackage[left=20mm,right=20mm,top=25mm,bottom=25mm]{geometry}
\usepackage{graphicx} 
\usepackage[usenames,dvipsnames]{xcolor}
\usepackage{xspace} 
\usepackage{bm} 
\usepackage{amsmath,amsfonts,amssymb,amsthm}
\usepackage[utf8]{inputenc} 

\xdefinecolor{mylinkcolor}{rgb}{0.0, 0.53, 0.74}
\usepackage[
	bookmarksnumbered, bookmarksopen, bookmarksopenlevel=2,
	breaklinks=true,
	colorlinks=true, filecolor=mylinkcolor, citecolor=mylinkcolor,
	linkcolor=mylinkcolor, urlcolor=mylinkcolor, menucolor=mylinkcolor,
]{hyperref}

\allowdisplaybreaks

\newcommand{\icasu}{\affiliation{Illinois Center for Advanced Studies of the Universe \& \\ Department of Physics University of Illinois at Urbana-Champaign.}}
\newcommand{\bex}{\bm{e}_x}
\newcommand{\bey}{\bm{e}_y}
\newcommand{\bez}{\bm{e}_z}
\newcommand{\bN}{\bm{N}}
\newcommand{\bP}{\bm{P}}
\newcommand{\bQ}{\bm{Q}}
\newcommand{\bL}{\bm{L}}
\newcommand{\bn}{\bm{n}}
\newcommand{\br}{\bm{r}}
\newcommand{\blambda}{\bm{\lambda}}
\newcommand{\GR}{\text{GR}}
\newcommand{\TT}{\text{TT}}
\newcommand{\ssGB}{\text{ssGB}}
\newcommand{\ST}{\text{ST}}
\newcommand{\lpar}{\left(}
\newcommand{\rpar}{\right)}

\begin{document}

\title{Theory-agnostic framework for inspiral tests of general relativity with higher-harmonic gravitational waves}

\hypersetup{pdftitle={TITLE}}

\author{Simone Mezzasoma}\email{simonem4@illinois.edu}\icasu
\author{Nicol\'as Yunes}\email{nyunes@illinois.edu}\icasu

\date{\today}

\begin{abstract}
Recent gravitational wave observations show evidence for the presence of higher harmonics, thus possibly indicating that these waves were generated in the inspiral of compact objects with asymmetric mass ratios.
Signals with higher harmonics contain a trove of information that can lead to a better estimation of system parameters and possibly to more stringent tests of general relativity. 
Gravitational wave model that include higher harmonics, however, have only been developed within general relativity, while models to test theory-agnostic deviations from general relativity have been purely based on the signal's dominant mode.
We here extend the parameterized post-Einsteinian framework to include the $\ell=2, 3$ and $4$ higher harmonics to first post-Newtonian order, therefore providing a ready-to-use Fourier-domain waveform model for tests of general relativity with higher harmonics.   
We find that the deformations to the higher harmonics of the Fourier phase can be easily mapped to the deformation of the dominant harmonic, while the deformations to the higher-harmonics of the Fourier amplitude in general cannot in a theory-agnostic way. 
Nonetheless, we develop a simple ansatz for the deformations of the waveform amplitude (through a re-scaling deformation of the time-domain amplitude) that both minimizes the number of independent amplitude deformations parameters and captures the predictions of all known modified theories to date.
\end{abstract}

\maketitle

\section{Introduction}
\label{sec:Introduction}

The advanced Laser Interferometer Gravitational-wave Observatory (aLIGO) \cite{LIGOScientific:2014pky} and advanced Virgo \cite{VIRGO:2014yos} have detected a plethora of events, some of which may have been produced by compact binaries with asymmetric mass ratios. For example, the most recent Gravitational-Wave Transient Catalog, GWTC-3, produced by the LIGO Scientific Collaboration (LSC) and the KAGRA \cite{KAGRA:2020tym} collaboration, contains 90 compact binary events, with a greater than $50\%$ probability of being of astrophysical origin~\cite{LIGOScientific:2021djp}. Of these, 15 sources have been estimated to have a (maximum-likelihood) mass ratio less than $1/2$. 

The quasi-circular inspiral of compact binaries with asymmetric mass ratios are interesting because they produce gravitational waves (GWs) that have power in many harmonics of the fundamental mode. From post-Newtonian (PN) theory, the GWs emitted in the quasi-circular inspirals of nearly equal mass compact binaries is dominated by the $(\ell , m)=(2,\pm 2)$ harmonic \cite{Blanchet:2008je}. Higher harmonics scale with powers of the mass difference, and thus, are important only for binaries with asymmetric mass ratios. The effect of these higher harmonics is to change the signal from effectively a sinusoid of monotonically-increasing phase and amplitude, to a superposition of waves with varying amplitudes and phases that lead to beats in the signal.

Precious information is contained in these higher modes, and if one does not model them, one may (i) introduce bias in the parameters extracted with a dominant-mode waveform model, and (ii) miss out on important physics that could have been extracted. For example, parameter estimation of GW170729 with a model that does not include higher harmonics can lead to a (systematic) mismodeling bias in the estimation of the individual masses of $\sim {\cal{O}}(10 \%)$ (see e.g.~Table I in~\cite{Chatziioannou:2019dsz}). Moreover, the inclusion of higher harmonics in the model can help constrain the mass ratio better by decreasing the confidence region by $\sim {\cal{O}}(15 \%)$ \cite{Chatziioannou:2019dsz}. Similar results were also found for the GW190814 event \cite{LIGOScientific:2020zkf}. Given this, one may also wonder whether tests of general relativity (GR) would be strengthened if one had a model to carry out such tests with higher harmonics.  

The disadvantages coming from neglecting the higher-harmonic content will only increase with the advent of third-generation detectors \cite{Divyajyoti:2021uty}, like the planned Einstein Telescope~\cite{Punturo:2010zz,Hild:2010id} and Cosmic Explorer~\cite{LIGODocumentT1500290,Dwyer:2014fpa,LIGOScientific:2016wof}. Upgraded detectors will be able to capture events with greater signal-to-noise ratio (SNR), which will make us more sensitive to biases in parameter estimation if we use theoretically incomplete models. Aside from improving the accuracy of parameter estimation, the introduction of higher harmonics helps break degeneracies, e.g. between inclination angle and luminosity distance \cite{Usman:2018imj}, which is particularly important in null tests of GR.

Many years of efforts have produced analytical waveform models that describe higher modes in GR.  The most recent ones, \texttt{PhenomPv3HM} \cite{Khan:2019kot} and \texttt{SEOBNRv4HM} \cite{Cotesta:2018fcv} (extensions of \texttt{PhenomPv3}~\cite{Khan:2018fmp} and \texttt{SEOBNRv4} respectively~\cite{Bohe:2016gbl}), have been employed by the LSC during the latest GW search \cite{LIGOScientific:2021sio}. Both these models reproduce the signal from spin-precessing black hole binaries and are key to correctly interpret events that show clear higher-multipole emission, as in the case of GW190412 \cite{LIGOScientific:2020stg}. Indeed, the analysis of this event with these models broke the degeneracy between the luminosity distance and the inclination angle, allowing for a measurement of both (see e.g.~Fig.~4 in \cite{LIGOScientific:2020stg}). 

Despite the success of such refined models in GR, not much effort has been put in developing a theory-agnostic beyond-quadrupole model that can be used to perform parametric tests of GR. As a result, the current Fourier-domain parameterized waveform models use only the dominant mode. A notable example is the parameterized Post-Einsteinian (ppE) model~\cite{Yunes:2009ke,Yunes:2010qb,Cornish:2011ys,Chatziioannou:2012rf,Loutrel:2014vja,Tahura:2018zuq,Nair:2020ggs,Tahura:2019dgr,Carson:2019kkh}, which incorporates generic beyond-GR deformations in both the amplitude and the phase, and has been used to successfully constrain modified theories of gravity since the early GW events (see e.g. \cite{Yunes:2016jcc}) through the TIGER implementation of the LSC~\cite{Agathos:2013upa,Li:2011vx,Meidam:2017dgf,LIGOScientific:2016lio,LIGOScientific:2018dkp,LIGOScientific:2019fpa,LIGOScientific:2020tif,LIGOScientific:2021sio}.

In this paper we tackle the task of extending the orignal ppE model to higher harmonics by proposing a  ready-to-use waveform template in the frequency domain for non-spinning circular compact binaries that includes the harmonics $\ell =2, 3,$ and $4$. Focusing on the inspiral, we achieve this by introducing beyond-GR perturbative corrections both in the time evolution of the orbital frequency and in the time-domain amplitude of each GW mode. Under the stationary phase approximation (SPA), each Fourier-domain harmonic inherits a ppE-like deformation that can be be uniquely tied to the orbital dynamics and the GW emission channels of the modified theory.

Our preliminary analysis shows that the ppE phase can be easily extended to all harmonics and the higher-harmonic contributions are simple rescalings of the dominant-mode harmonic. The ppE amplitude corrections that arise in higher harmonics, however, cannot be mapped only to the dominant-mode amplitude in general. This is because the ppE phase parameter of each harmonic is uniquely determined by the time evolution of the orbital frequency, accurate to leading-PN order in the non-GR deformation. The ppE amplitude corrections, on the other hand, depend in general on not just the orbital trajectories, but also on the perturbed field equations for the GW metric perturbation. Nonetheless, we show that a simple ansatz for the metric perturbation minimizes the number of independent ppE amplitude parameters while simultaneously capturing the predictions of all modified theories of gravity at 1PN-order known to date.

It remains to be seen if this simple ppE extension is sufficient to cover other modified theories, but in the meantime, the basic tool developed here can still be used to constrain theory-agnostic deviations from GR with GWs that contain higher-harmonics. In particular, our ppE extension can be used to assess the gains in the strength of GW tests of GR due to the inclusion of higher harmonics. If this ppE extension significantly tightens the current bounds on the coupling constants of modified theories, then one could further refine the extension by (i) including degeneracy-breaking physics (e.g. elliptical orbits and spin precession) and (ii) putting forward a more sophisticated ansatz that maps to a broader set of theories (once the predictions of the latter have been worked out). 

The breakdown of the paper is as follows. In Sec.~\ref{sec:HigherHarmonicsGR}, we review the higher harmonic decomposition of GWs from inspiraling non-spinning binaries withing GR. Specifically, we lay out the 1.5PN-order GR results that will serve as the basis for the subsequent ppE construction. In Sec.~\ref{sec:ppEnutshell}, we re-derive the original ppE waveform and we discuss how the issue of computing the dominant-mode amplitude correction has been approached in previous work.  Section \ref{sec:NewppE} contains the new higher-harmonic ppE waveform model and its minimal version, which is shown to properly represent the predictions of two known theories. In Sec.~\ref{sec:Conclusions}, we summarize our findings and suggest future avenues for future work. Lastly, App.~\ref{app:GR_waveform_amplitudes} collects the coefficients that define the 1.5PN-order GR waveform and App.~\ref{app:Spin2_Spherical_Harmonics} summarizes the definition of spin-weighted $s=2$ spherical harmonics. We adopt geometric units, thus setting $G=1=c$ throughout.

%%%%%%%%%%%%%%%%%%%%%%%%%%%%%%%%%%%%%%%%%%%%%%%%%%%%%%%%%%%%%%%%%%%%%%%%%%%
\section{Higher harmonics in GR}
\label{sec:HigherHarmonicsGR}

In this section, we outline the known harmonic content of the GW waveform from an inspiraling, non-spinning compact binary, up to 1.5PN-order. The time domain harmonics are then Fourier-transformed under the stationary-phase approximation and the leading-PN order result of each harmonic is presented. 

%----------------------------------------------------------------------------
\subsection{Full waveform in GR}
\label{sec:Full_waveform}
Let us consider a GW produced in GR by an isolated slow-moving inspiraling binary source. First, we define the basis used to decompose the GW strain tensor $h_{ij}$ and the vectors that describe the dynamics of the binary system. Following the conventions in \cite{Blanchet:2008je}, we construct the source orthonormal basis $\lbrace \bex ,\bey ,\bez \rbrace$ where $\bez$ is aligned with the orbital angular momentum $\bL$ of the binary. Since we restrict our analysis to the quasi-circular motion of non-spinning or spin-aligned binaries, the vector $\bL$ is assumed to be fixed (see e.g. \cite{Arun:2008kb} for an analysis on the time evolution). The individual positions $\br_1$ and  $\br_2$ of the binary components relative to the center of mass are mapped to the separation vector $\br \equiv \br_1 - \br_2$ in the center-of-mass frame by requiring that the source dipole moment, conserved under the equations of motion, is zero \cite{Blanchet:2002mb}. Because the motion is constrained to the orbital plane, we can equivalently describe the system with the separation distance $r\equiv|\br|$ and the orbital phase $\phi$. Then, the unit vector 
\begin{equation}
	\bn\equiv\br/r =\cos\phi\,\bex +\sin\phi\,\bey ,
\end{equation}
is used to build the co-rotating frame $\lbrace \bn ,\blambda ,\bez \rbrace$, with $\blambda\equiv\bez \times \bn =-\sin\phi\,\bex+\cos\phi\,\bey $.

The detector lies far from the source at a distance $D$ in the direction of the unit vector 
\begin{equation}
	\bN \equiv \sin\iota \cos\varphi \, \bex + \sin\iota \sin\varphi \, \bey+\cos\iota \, \bez ,
\end{equation}
with inclination angle $\iota\in \left[0,\pi\right]$ and azimuthal angle $\varphi\in [0,2\pi)$. In a radiative coordinate system $X^\mu=(T,X^i)$, with $D=(X^iX^j \delta_{ij})^{1/2}$, the metric at the detector location takes the form
\begin{equation}
	g_{\mu\nu} = \eta_{\mu\nu} + h_{\mu\nu} +\mathcal{O}(h^2) ,
\end{equation} 
where $\eta_{\mu\nu}=\text{diag}(-,+,+,+)$ is the Minkowski metric. We work in the transverse-traceless (TT) gauge, so the spatial components of the gravitational potentials $h^\TT_{\mu\nu}$, carrying the GW from a PN source, can be written as a radiative multipolar series (e.g. Eq.~(2.1) in \cite{Blanchet:2008je}). The leading term in $1/D$ of $h^\TT_{ij}$ is then used to construct the plus and cross waveform polarizations
\begin{align}\label{h_plus_cross}
	h_+ &\equiv \frac{1}{2}(P^iP^j-Q^iQ^j)h^\TT_{ij} ,\\
	\label{h_cross}
	h_\times &\equiv \frac{1}{2}(P^iQ^j+Q^iP^j)h^\TT_{ij} .
\end{align}
Here the vectors $\bP$ and $\bQ$ are part of the polarization triad $\lbrace \bN, \bP,\bQ\rbrace$, and are defined through
\begin{align}
	\bP \equiv& \bN\times \bez=\sin\varphi\, \bex - \cos\varphi\, \bey,\\
	\bQ\equiv&\bN\times \bP =\cos\iota\,\cos\varphi\,\bex+\cos\iota\sin\varphi\,\bey-\sin\iota\,\bez .
\end{align}

For a binary system composed of compact objects of masses $m_1$ and $m_2$ in a quasi-circular orbit, Eqs.~\eqref{h_plus_cross} and~\eqref{h_cross} produce the PN series
\begin{equation}\label{h_plus_cross_PN}
	h_{+,\times} =\frac{2\mu}{D}x\sum_{j=0}^\infty x^{j/2} H_{+,\times}^{(j/2)} ,
\end{equation}
where $\mu =(m_1 m_2)/M$ is the reduced mass, $M=m_1+m_2$ is the total mass, and 
\begin{equation}
    x(t)\equiv (M \dot{\phi}(t))^{2/3}
\end{equation}
is the PN expansion parameter. The coefficients $H_{+,\times}^{(j/2)}$ are functions of the orbital phase and the inclination angle, which are presented up to 3PN order for example in in Sec.~VIII of \cite{Blanchet:2008je} after replacing 
\begin{equation}
	\psi \rightarrow \psi - \varphi +\pi/2 ,
\end{equation}
where 
\begin{equation}
	\psi \equiv \phi - 3 x^{3/2}\left(1- \frac{\nu}{2}x\right)\ln(x/x_0) ,
\end{equation}
is the tail-distorted phase variable \cite{Blanchet:1993ec} and $x_0$ is a gauge-dependent arbitrary constant. This auxiliary phase variable is introduced to conveniently recast logarithmic terms present in the amplitude into a phase modulation, forcing those terms to appear in the waveform only as 3PN (rather than 1.5PN \cite{Kidder:2007rt}) amplitude corrections.

\subsection{Waveform harmonic decomposition}
\label{sec:Harmonic_decomposition}
By inspecting the coefficients $H_{+,\times}^{(j/2)}$ we see that, up to 1.5PN order included ($j \leq 3$), the time-domain gravitational waveform in Eq.~\eqref{h_plus_cross_PN} can be rearranged as a superposition of modes in the form
\begin{equation}\label{h_plus_cross_modes}
	h_{+,\times} = \sum_{n=1}^5 A_{+,\times}^{(n)}(x) \cos\Phi_{+,\times}^{(n)} + \mathcal{O}(x^3) ,
\end{equation}
where $\Phi_+^{(n)} \equiv n(\psi - \varphi + \pi/2)$, $\Phi_\times^{(n)} \equiv \Phi_+^{(n)} - \pi/2$, and the amplitudes 
\begin{equation}\label{PN_amplitdue_decomposition}
	A^{(n)}_{+,\times} (x) = -\frac{2\mu}{D}x \sum_{j=0}^3 a^{(n,j)}_{+,\times} x^{j/2} ,
\end{equation}
are given by a PN series whose coefficients are listed in Appendix \ref{app:GR_waveform_amplitudes}. Note that we have dropped the non-linear $n=0$ memory terms \cite{Favata:2008yd}  in Eq.~\eqref{h_plus_cross_modes} as our target detectors are limited by a positive cutoff frequency \cite{Blanchet:2013haa}. 

For data analysis applications, the analytic expression in Eq.~\eqref{h_plus_cross_modes}, only valid in the early inspiral stage, has to be matched to numerical relativity simulations that cover the merger-ringdown phase (see e.g. \cite{Jani:2016wkt,Boyle:2019kee}). To make contact with numerical results, it is routine to define the complex combination
\begin{equation}
	\text{h}(t)  \equiv h_+ - i \, h_\times ,
\end{equation}
and extract the $(\iota , \varphi)$ angular dependence by projecting it on the basis of spin-weighted spherical harmonics $ Y_{-s}^{\ell m}(\iota , \varphi) $ (see Appendix~\ref{app:Spin2_Spherical_Harmonics}  for definitions) with spin $s=2$
\begin{equation}
	\text{h}(t) = \sum_{\ell =2}^\infty \sum_{m=-\ell}^{\ell}h_{\ell m}(t)\, Y_{-2}^{\ell m}(\iota , \varphi) .
\end{equation}
The harmonic coefficients are obtained by integrating
\begin{equation}\label{h_ell_m_integral}
h_{\ell m}(t) = \int_{-1}^1 d \cos\iota \int_0^{2\pi}d\varphi \,\,\text{h}(t) \left( Y_{-2}^{\ell m}(\iota , \varphi)\right)^* ,
\end{equation}
and, for non-precessing binaries, they satisfy the equatorial symmetry
\begin{equation}\label{h_equatorial_symmetry}
	h_{\ell, -m}(t) = (-1)^\ell h_{\ell m}^*(t) ,
\end{equation}
which allows us to recover the components with $m<0$ from those with $m>0$. Henceforth, we will only focus on the positive values of $m$. Performing the integration in Eq.~\eqref{h_ell_m_integral} shows that $h_{\ell m}(t)$ can be written as
\begin{equation}\label{h_ell_m_time_domain}
	h_{\ell m}(t) =	\mathcal{A}_{\ell m}(x) \, e^{-i m \psi} ,
\end{equation}
where
\begin{align}\label{mathcalA_general}
    \mathcal{A}_{\ell m}(x) &\equiv \frac{\sqrt{(2\ell +1)\pi}}{2} (-i)^m \times \\ \nonumber
    & \times \int_{-1}^1 d\cos\iota \left( A_+^{(m)}+A_\times^{(m)}\right) d_2^{\ell m}(\cos\iota) ,
\end{align}
and the functions $d_2^{\ell m}$ are defined in Eq.~\eqref{Spin2_Spherical_Harmonics_d}.

With the  harmonic content in Eq.~\eqref{PN_amplitdue_decomposition}, which includes the $\ell = 2,3,4$ modes, Eq.~\eqref{mathcalA_general} takes the form
\begin{equation}\label{mathcalA_PN}
	\mathcal{A}_{\ell m}(x) = \frac{2\mu}{D}\sqrt{\frac{16\pi}{5}}x\, \mathcal{H}_{\ell m}(x) ,
\end{equation}
and the leading PN contribution (for the higher PN order expressions, see~\cite{Blanchet:2008je}) of each frequency-dependent amplitude $\mathcal{H}_{\ell m}$ is 
\begin{subequations}\label{mathcalH_ell_m}
\begin{align}
	\mathcal{H}_{2 1} &= \frac{i}{3}x^{1/2} \Delta , \\
	\mathcal{H}_{2 2} &= 1 ,\\
	\mathcal{H}_{3 1} &= \frac{i  }{12 \sqrt{14}}x^{1/2}  \Delta ,\\
	\mathcal{H}_{3 2} &= \frac{1}{3} \sqrt{\frac{5}{7}} (1-3 \nu) x ,\\
	\mathcal{H}_{3 3} &=-\frac{3}{4} i \sqrt{\frac{15}{14}} x^{1/2}\Delta ,\\
	\mathcal{H}_{4 1} &=\frac{i}{84 \sqrt{10}}(1-2 \nu)x^{3/2}\Delta ,\\
	\mathcal{H}_{4 2} &=\frac{ \sqrt{5}}{63} (1-3 \nu) x ,\\
	\mathcal{H}_{4 3} &=-\frac{9 i  }{4 \sqrt{70}}(1-2 \nu )x^{3/2} \Delta ,\\
	\mathcal{H}_{4 4} &=-\frac{8}{9} \sqrt{\frac{5}{7}} (1-3 \nu) x ,
\end{align}
\end{subequations}
where $\Delta \equiv \text{sgn}(m_1-m_2) \sqrt{1-4\nu}$, and $\nu = \mu/M$ is the symmetric mass ratio.

We note that each mode in Eq.~\eqref{h_ell_m_time_domain} involves a slow-varying amplitude multiplied by  an oscillating phase, i.e. 
\begin{equation}
	\left|\frac{d }{dt}\ln \mathcal{A}_{\ell m}\right|\ll |\dot{\psi}| ,
\end{equation} 
and assuming $|\ddot{\psi}|\ll\dot{\psi}^2$, we can obtain an analytic expression of its Fourier transform
\begin{equation}
	\tilde{h}_{\ell m}(f) \equiv \int_{-\infty}^\infty h_{\ell m}(t)\, e^{2\pi i f t}dt ,
\end{equation}
under the SPA \cite{Cutler:1994ys,Droz:1999qx,Yunes:2009yz} with
\begin{equation}\label{h_lm_SPA_GR}
		\tilde{h}_{\ell m}(f) = \left(\frac{2\pi}{m\ddot{\psi}(\bar{t}_m )}\right)^{1/2} \mathcal{A}_{\ell m}(\bar{x}_m)\, e^{-i \left( m \Psi(\bar{x}_m) + \pi/4 \right)} .
\end{equation}
In the last expression, valid for $m>0$ and $f>0$, the stationary time $\bar{t}_m$ is implicitly defined by
\begin{equation}
	\dot{\psi}(\bar{t}_m) = \frac{2\pi f}{m} ,
\end{equation}
or equivalently by
\begin{equation}\label{x_stationary}
	\bar{x}_m \equiv x(\bar{t}_m)=\left(\frac{2\pi M f}{m}\right)^{2/3} +\mathcal{O}[(Mf)^{11/3}] .
\end{equation}
The phase $\Psi(\bar{x}_m)$ appearing in Eq.~\eqref{h_lm_SPA_GR} is defined as
\begin{align}\label{Fourier_phase}
	\Psi(\bar{x}_m)&\equiv \psi(\bar{t}_m) -\frac{2\pi f}{m}\bar{t}_m -\psi_c + \frac{2\pi f}{m}t_c \\
	&=\frac{1}{M}\Big[ \int^{\bar{x}_m}  \frac{x^{3/2}}{\dot{x}}dx - \bar{x}_m^{3/2}\int^{\bar{x}_m}  \frac{dx}{\dot{x}} \Big] + \\ \nonumber
	&-\psi_c + \frac{\bar{x}_m^{3/2}}{M} t_c ,
\end{align}
where $\psi_c,t_c$ are integration constants,
\begin{equation}
	\dot{x} \equiv \frac{dx}{dt} = \frac{64}{5}\frac{\nu^2}{\mu}x^5 + \mathcal{O}(x^6),
\end{equation}
and we have neglected sub-leading PN terms arising from the mismatch between $\psi$ and $\phi$.

To leading PN order in both amplitude and phase, Eq.~\eqref{h_lm_SPA_GR} for non-spinning binaries in a quasi-circular orbit reads \cite{Mishra:2016whh}
\begin{align}\label{h_GR_harmomics}
	\tilde{h}_{\ell m}^\GR(f) &\equiv \frac{\pi M^2}{D}\sqrt{\frac{2\nu}{3}}\bar{x}_m^{-7/4} \left(\frac{2}{m}\right)^{1/2}\mathcal{H}_{\ell m}(\bar{x}_m)\times\\ \nonumber
	&\times \exp \left[i m   \Big(\frac{3}{256 \nu}\bar{x}_m^{-5/2} + \psi_c \Big) - i\pi/4\right] ,
\end{align}
with $\bar{x}_m$ defined by Eq.~\eqref{x_stationary} and the frequency-dependent amplitudes $\mathcal{H}_{\ell m}(x)$ given in Eqs.~\eqref{mathcalH_ell_m}.

%%%%%%%%%%%%%%%%%%%%%%%%%%%%%%%%%%%%%%%%%%%%%%%%%%%%%%%%%%%%%%%%%%%%%%%%%%%
\section{ppE in a nutshell}
\label{sec:ppEnutshell}

In this section, we review the building blocks of the dominant-mode ppE waveform model. After defining the small coupling limit, we illustrate how the ppE waveform model can be obtained from a radial perturbation of the two-body Lagrangian. We highlight the known mapping between ppE parameters and beyond-GR deformations of the orbital dynamics, emphasizing how the amplitude correction requires the choice of an ansatz for the metric perturbation.

%----------------------------------------------------------------------------
\subsection{The ABC of ppE}
\label{subsec:ABC-ppE}
The ppE framework, first proposed in \cite{Yunes:2009ke}, describes deviations beyond GR encoded in GWs from compact binaries. The strategy behind the ppE formalism is to work in the perturbative regime of a chosen metric theory of gravity in order to have analytic control of the waveform with just a few deformation parameters. The metric theory must admit a well defined and continuous limit to GR in the weak-field and low-velocity regime, and it is required to have an observable non-GR effect in the strong-field regime.

Considering only the GR polarization content, the Fourier transform of the response function due to the GW impinging on a two-arm $90^\circ$-interferometer is 
\begin{equation}\label{strain}
	\tilde{h}_r(f)\equiv F_+\tilde{h}_+ (f) +F_\times \tilde{h}_\times (f),
\end{equation}
where $F_{+,\times}$ are the detector beam-pattern coefficients \cite{Finn:1992xs,Cutler:1997ta}
which depend on the source location on the sky $(\theta_S,\phi_S)$ and the polarization angle $\psi_S$. These coefficients describe the relative orientation of the source with respect to the detector and we have assumed that the signal remains in band for a short enough time to regard the angles $(\theta_S,\phi_S,\psi_S)$ as constant. This justifies taking the Fourier transform of the metric perturbation before combining it to form the response function.

During the early inspiral phase of the binary, the simplest ppE template of the sky-averaged Fourier domain response function is
\begin{equation}\label{simplestppE}
	\tilde{h}_r(f)=\tilde{h}_{r,\GR}(f)(1+\alpha u^a) e^{i \beta u^b} ,
\end{equation} 
with $u \equiv (\pi \mathcal{M}f)^{1/3}$, and the leading-PN order GR response function (higher-order phase corrections can be found e.g. in Eq.~(3.18) of \cite{Buonanno:2009zt}) is
\begin{equation}\label{leadingTaylorF2}
	\tilde{h}_{r,\GR}(f)\equiv \sqrt{\frac{\pi}{30}}\frac{\mathcal{M}^2}{D}u^{-7/2} \exp \lpar i \frac{3}{128} u^{-5}+\mathcal{O}(u^{-3})\rpar ,
\end{equation} 
where $\mathcal{M}=(m_1 m_2)^{3/5}/(m_1+m_2)^{1/5}$ is the chirp mass ($m_1$ and $m_2$ denoting the component rest masses), $D$ is the distance from the source, and $f$ is the measured GW frequency. 
Each theory corresponds to a particular set of ppE parameters $\lbrace \alpha, a,\beta, b\rbrace$, of which $\alpha$ and $\beta$ measure the magnitude of the corrections and are functions of the coupling constant of the theory as well as the intrinsic masses and spins of the binary. The exponents $a$ and $b$ take (positive or negative) integer values and control the PN order at which these corrections enter the amplitude and the phase, respectively. The gravitational waveform computed from many alternative theories of gravity in the small coupling limit has been shown to produce Eq.~\eqref{simplestppE}, to leading PN order \cite{Tahura:2018zuq,Tahura:2019dgr,Chatziioannou:2012rf}.

Let us now define the small-coupling limit. 
When expanding the amplitude or the phase of the Fourier-domain response function 
$\tilde{h}_r(f)$ in terms of two independent variables, such as the dimensionless coupling $\zeta$ of the theory and the velocity of the system $v\propto u^3$, we need to choose the order in which we perform the expansion as the two operations may not commute.
To illustrate this, consider the simple function
\begin{equation}\label{series_example}
	d(\zeta,v)=\frac{1}{\zeta+v},
\end{equation}
which is singular along the line $\zeta = -v$ and in particular at the origin. Expanding first in powers of $\zeta$ and then in $v$ generates 
\begin{equation}
	d_\text{I}(\zeta,v)=\sum_{n=0}^{\infty}\frac{(-1)^n}{v^{n+1}}\zeta^n ,
\end{equation} 
which converges to $d(\zeta,v)$ in the region $|\zeta|<|v|$. Alternatively, one could expand first in powers of $v$ and subsequently in $\zeta$, obtaining
\begin{equation}
	d_\text{II}(\zeta,v)=\sum_{n=0}^{\infty}\frac{(-1)^n}{\zeta^{n+1}}v^n ,
\end{equation}
which is valid in the complementary region $|v|<|\zeta|$. Clearly the two representations $d_\text{I}$ and $d_\text{II}$ are not equivalent and picking one over the other corresponds to restricting the parameter space that we wish to study.

In this paper, if $\tilde{h}_r(f)$ is singular around $(\zeta,v)=(0,0)$ and the standard Taylor series cannot be used to approximate the function in this region, we resort to a power series representation by expanding first to leading order in the coupling $\zeta$ and subsequently in the velocity $v$. This corresponds to choosing $d_\text{I}$ in the above example and it is often referred to as \textit{quadrupole-driven} case as opposed to \textit{dipole-driven} case \cite{Shiralilou:2021mfl,Khalil:2018aaj}.
This ensures that if the GR limit exists, taking $\zeta\rightarrow 0$ in the expression of  $\tilde{h}_r(f)$ yields a GR PN approximant, such as \texttt{TaylorF2} \cite{Buonanno:2009zt}, whose leading contribution is in Eq.~\eqref{leadingTaylorF2}.

%----------------------------------------------------------------------------
\subsection{Waveform derivation}

\subsubsection{Time domain waveform model}
\label{sssec:Time_domain_waveform_model}
Since the original ppE formulation was developed for leading-order~PN calculations, only modifications to the dominant $(\ell,m)=(2,\pm2)$ mode could be probed. In this paper, we will extend this formulation to 1PN order, so that higher harmonics can also be probed. But before we do so, let us review the procedure to obtain Eq.~\eqref{simplestppE} for a non-precessing binary in quasi-circular equatorial orbits.

We start with a two-body PN Lagrangian
\begin{equation}
	L=L_0+\zeta L_1\label{Lagrangian} ,
\end{equation}
in the center-of-mass spherical coordinates $(r,\phi)$, where $r(t)$ is the relative separation and $\phi(t)$ is the orbital phase of the binary. The quantity $L_0$ is a two-body Lagrangian obtained from GR, for instance through the Fokker action method \cite{Fokker:1929,Bernard:2015njp}. For leading-PN-order calculations, it is sufficient to keep $L_0$ to Newtonian order, i.e., $L_0=\mu (\dot{r}^2+r^2\dot{\phi}^2)/2+M\mu /r$, where we recall that $M=m_1+m_2$ is the total mass and $\mu=(m_1m_2)/M$ is the reduced mass. The quantity $L_1$ encodes an effective, non-GR Lagrangian modification multiplied by a dimensionless parameter $\zeta$, which is linked to the coupling constant of the theory.
We assume this parameter to be small and use it as an expansion variable following the prescription outlined in Sec.~\ref{subsec:ABC-ppE}.

Like $L_0$, $L_1$ is generated as a PN series, for example, using the Fokker formalism applied to modified theories of gravity \cite{Damour:1995kt, Bernard:2018hta,Julie:2019sab,Khalil:2018aaj}. One could also follow theory-agnostic methods, such as the modified Einstein–Infeld–Hoffmann framework \cite{Will:2018ont}, to first parameterize the two-body Lagrangian to a given PN order and then retain only the leading terms that survive after the small coupling expansion. However, we will not employ this kind of approach here as it would lead to involved equations of motion, going beyond the scope of this work.

To exemplify, we consider deformations of the form 
\begin{equation}\label{L_1}
	L_1 =\mu\lpar\frac{M}{r}\rpar^w,
\end{equation}
although we stress that for a consistent analysis such modifications should include a dependence on $\dot{\phi}$ and $\dot{r}$. 
In fact, velocity-dependent corrections in $L_1$ may arise from genuine modifications of the kinetic term due to beyond-GR effects \cite{Loutrel:2014vja}, but also from our choice of center-of-mass coordinates \cite{Blanchet:2002mb,Mirshekari:2013vb,Lang:2013fna}. 

The exponent $w$ in Eq.~\eqref{L_1} determines the leading PN order characteristic properties of the theory. In what follows we consider $w>1$ to exclude degeneracy with the Newtonian potential; the case $w=1$ can be reabsorbed into $L_0$ after a redefinition of the total mass $M$ or the gravitational constant $G$. For simplicity, we allow $w$ to take only integer values to preserve the time-reversal symmetry of the conservative sector.  

We proceed by deriving the Euler-Lagrange equations from Eq.~\eqref{Lagrangian} and imposing the condition of circular orbits 
\begin{equation}\label{circular}
	\dot{r}=0=\ddot{r}.
\end{equation}
This leads to a modified Kepler's third law of the form
\begin{equation}\label{kepler}
	\frac{M}{r}=x(1+ \zeta \lambda x^p) ,
\end{equation}
where $\lambda\equiv -w/3$, $p \equiv w-1$, and the PN frequency parameter is $x = (M \dot{\phi})^{2/3} $.

Notice that $M$ here is the active gravitational mass and we have set the gravitational constant $G$ to unity. All of this allowed because the leading order modification enters at higher than Newtonian order; if these weren't the case, then the active mass or the gravitational constant would have to be renormalized.

Legendre-transforming Eq.~\eqref{Lagrangian} leads to the Hamiltonian of the system
\begin{equation}
	H=H_0 +\zeta H_1 \label{hamiltonian},
\end{equation}
with $H_0 = \mu(\dot{r}^2+r^2\dot{\phi}^2)/2-M\mu /r$ and $H_1 = -L_1$, where one can think of the velocities as implicit functions of the momenta $p_r=\partial L /\partial \dot{r}$, $p_\phi=\partial L /\partial \dot{\phi}$.
We then use the equations of motion in Eqs.~\eqref{circular}-\eqref{kepler} inside the Hamiltonian of Eq.~\eqref{hamiltonian} to obtain the binding energy as a function of $x$, namely
\begin{equation}\label{binding_E}
	E(x)=- \frac{\mu x}{2}(1+\zeta A x^p),
\end{equation}
with $A\equiv -2(2w-3)/3$. In this expression we have discarded $\mathcal{O}(\zeta^2)$ and $\mathcal{O}(\zeta x^{p+2})$ terms for consistency, but higher PN order terms can in principle be included straightforwardly. 

Independently from the conservative dynamics provided by Eq.~\eqref{Lagrangian} we must now prescribe the energy flux $\mathcal{F}$ due to all radiating fields present in the theory (other fluxes are not relevant to quasi-circular orbits at leading PN order). For example, for scalar-tensor theories \cite{Damour:1992we,Lang:2014osa}, the leading PN flux $\mathcal{F}=\mathcal{F}_T$+$\mathcal{F}_S$ is comprised of a tensorial GW contribution $\mathcal{F}_T=(32\pi)^{-1}\int (\dot{h}_{ij})^2 D^2d\Omega$ and a scalar contribution $\mathcal{F}_S=(4\pi)^{-1}\int ( \dot{\Psi})^2 D^2 d\Omega$, where $\Psi$ denotes the scalar field and here the overdot is the derivative with respect to retarded time $t_\text{ret}\equiv t-D$. In general, the fluxes are constructed from derivatives of the fields, which can be expanded in terms of radiative multipole moments in the far-away radiation zone, and which are, in turn, given in terms of the source multipole moments through asymptotic matching. Evaluating the energy flux through the equations of motion in Eqs.~\eqref{circular}-\eqref{kepler}, trading any $r$ and $\dot{\phi}$ dependence for $x$, and linearizing in $\zeta$ and to leading PN order in $x$, one typically obtains a frequency-dependent expression of the form
\begin{equation}\label{flux}
	\mathcal{F}(x)=\frac{32}{5}\nu^2 x^5(1+\zeta B x^q) ,
\end{equation}
for some dimensionless coefficient $B \in \mathbb{R}$ that will depend on the intrinsic parameters of the binary. Ignoring the presence of non-GR hereditary tails, which in GR enter at 1.5PN order \cite{Blanchet:2013haa}, we take the exponent $q$ to be a non-zero (possibly negative) integer.

With the conservative and the dissipative sectors parameterized in terms of GR deformations, we can now obtain the time evolution of the frequency through the balance equation
\begin{equation}\label{inverse_x_dot}
	\dot{x}^{-1}(x)\equiv\frac{dt}{dx} =-E'(x)/\mathcal{F}(x)\,.
\end{equation} 
Expanding this equation in both $\zeta$ and $x$, we obtain
\begin{equation}\label{oneoverxdot_ppE}
\dot{x}^{-1}(x)=\frac{5}{64}\frac{\mu}{\nu^2}x^{-5}(1+\zeta C x^k) ,
\end{equation}
where 
\begin{equation}\label{oneoverxdot_ppE_k_C}
	k\equiv \min (p,q),\qquad C\equiv A(p+1)\delta_{k,p}-B\delta_{k,q},
\end{equation}
and $\delta_{p,q}$ is the Kronecker symbol. In Eq.~\eqref{oneoverxdot_ppE} we have consistently discarded $\mathcal{O}(\zeta^2)$ and $\mathcal{O}(\zeta x^{k-4})$ terms.

In order to proceed, we must now select a parameterization for the deformations of the metric perturbation $h^\TT_{ij}$ in the TT gauge, requiring it to be compatible with the multipole expansion used to produce the energy flux in Eq.~\eqref{flux}. An exact description of the GW waveform can only be achieved after selecting a particular theory and solving for its polarization content \cite{Chatziioannou:2012rf}, as done e.g. by integrating directly the field equations \cite{Will:1996zj,Pati:2002ux}.  Without specifying the theory one can only propose an ansatz for the plus and cross  waveform polarizations. Commonly employed for leading PN-order estimations \cite{Tahura:2018zuq,Alexander:2018qzg} is the quadrupole formula (see e.g. $\S 1.2$  \cite{Blanchet:2013haa} for the formal definition), which in the time domain and for a quasi-circular binary implies
\begin{align}\label{h_generic_quadrupole}
	h_{+,\times}(x)&=A_{+,\times}(x)\cos \Phi_{+,\times} ,
\end{align}
with $\Phi_+ \equiv 2\phi $, $\Phi_\times \equiv \Phi_+ -\pi/2$. The amplitudes are taken to be \cite{Tahura:2018zuq}
\begin{equation}\label{ansatz_ampl}
	A_{+,\times}\equiv \frac{2\mu}{D}Q_{+,\times} r^2 \dot{\phi}^2 ,
\end{equation}
and are reduced to functions of $x$ by means of Eq.~\eqref{kepler}
\begin{equation}
	A_{+,\times}(x)= \frac{2\mu }{D}Q_{+,\times}x(1+2\zeta \lambda x^p).
\end{equation}
Here $Q_+ \equiv -(1+\cos^2\iota)$ and $Q_\times \equiv -2\cos\iota$ are functions of the inclination angle $\iota$. 

We see that the modified Lagrangian in Eq.~\eqref{L_1} stretches the amplitude by a frequency-dependent term which is of the same PN order as the correction affecting the binding energy in Eq.~\eqref{binding_E}. Note, however, that Eq.~\eqref{ansatz_ampl} is constructed following the quadrupole formula of GR, and it relies on the Newtonian equations of motion to eliminate the accelerations that arise from the second time derivative of the quadrupole moment. Thus, depending on our starting point in the parameterization of $	h_{+,\times}$, some effects of order $\mathcal{O}(\zeta x^{p+1})$ in the amplitude may be lost at this stage. To overcome this problem, we may generalize the amplitude deformation to
\begin{equation}\label{generalized_amplitdue}
	A_{+,\times}(x)\rightarrow A_{+,\times}^{\GR}(x)(1+ \zeta \,\Gamma_{+,\times} x^{\gamma_{+,\times}}),
\end{equation}
with $A_{+,\times}^{\GR}(x)=2\mu Q_{+,\times}x/D$ and where the coefficients and exponents $\lbrace\Gamma_{+,\times},\gamma_{+,\times}\rbrace$ parameterize the type of GR deformation. Given a particular theory, these new ppE coefficients can be read from the waveform, as was done e.g.~in \cite{Chatziioannou:2012rf}, after integrating the field equations consistently in the small coupling approximation. One point to note is that, at the level of Eq.~\eqref{generalized_amplitdue}, we can increase the PN accuracy of $A_{+,\times}^{\GR}(x)$ by replacing it with $A_{+,\times}^{(2)}(x)$ in Eq.~\eqref{PN_amplitdue_decomposition} at the cost of introducing mismodeled terms of order $\mathcal{O}(\zeta x^{\gamma_{+,\times}+3/2})$. Because $\zeta$ is parameterically small and the analysis is restricted to quadrupolar radiation, we expect these terms to have little impact on parameter estimation, while the presence of higher-PN GR terms helps to cover a longer segment of the inspiral. 

\subsubsection{Frequency domain waveform model}

The Fourier-transform of each waveform polarization in Eq.~\eqref{h_generic_quadrupole} is readily computed under the SPA. Denoting $x_f\equiv \bar{x}_2 = (\pi M f)^{2/3}$, we have
\begin{align}
	\tilde{h}_{+,\times}(f)&\equiv \int_{-\infty}^{\infty}h_{+,\times}(x) e^{2\pi i f t}dt \\
	&= \tilde{A}_{+,\times}(x_f)\exp \left[ -i( 2\Psi_{+,\times}(x_f)+\pi/4)\right] ,
\end{align}
where
 \begin{align}
 	\tilde{A}_{+,\times}(x_f)&\equiv \sqrt{\frac{\pi M }{6}}x_f^{-1/4}\sqrt{\dot{x}^{-1}(x_f)} A_{+,\times}(x_f) \label{Ampl_tilde} ,
 \end{align}
and
\begin{align}
    \Psi_+(x_f)&\equiv \Psi(x_f)\label{Psi_tilde} ,\\
    \Psi_\times(x_f)&\equiv \Psi_+(x_f)-\pi/4 ,
 \end{align}
with $\Psi(x_f)$ given by Eq.~\eqref{Fourier_phase}, which now contains the modified $\dot{x}^{-1}$ in Eq.~\eqref{oneoverxdot_ppE}.

Now taking the small coupling limit of Eqs.~\eqref{Ampl_tilde}-\eqref{Psi_tilde} leads to
\begin{equation}
	\tilde{A}_{+,\times}(x_f)=\frac{M^2}{4D}\sqrt{\frac{5\pi\nu}{6}}Q_{+,\times}  x_f^{-7/4}\left( 1+\zeta H x_f^k \right)\label{SPAamplitude} ,
\end{equation}
with 
\begin{equation}
	H\equiv \frac{C}{2}(\delta_{k,p}+\delta_{k,q}-\delta_{p,q})+2\lambda \delta_{k,p} ,
\end{equation}
and
\begin{equation}\label{SPAphase}
	\Psi_+(x_f)=-\frac{3}{256\nu}x_f^{-5/2}\left( 1+\zeta W x_f^k\right) ,
\end{equation}
with
\begin{equation}\label{W}
W\equiv  \frac{20}{(k-4)(2k-5)}C .
\end{equation}
In Eq.~\eqref{W} we excluded the case $k=5/2$ as it would only occur when non-GR tail-like terms are included in the flux. We further assumed $k\neq 4$ for simplicity, as among the theories that have been studied (e.g. see Table III of \cite{Yunes:2016jcc}) this case only appears in a specific Gravitational Standard Model Extension \cite{Kostelecky:2016kfm}.

Inserting Eqs.~\eqref{SPAamplitude}-\eqref{SPAphase} into \eqref{strain} we obtain
\begin{equation}
	\tilde{h}_r(f)=Q \frac{M^2}{4D}\sqrt{\frac{5\pi\nu}{6}}  x_f^{-7/4}\left( 1+\zeta H x_f^k \right) \exp(-i 2\Psi_+(x_f)) ,
\end{equation}
up to an overall constant phase $e^{i\Phi}$ with
\begin{equation}
    \Phi\equiv \text{arctan2}(F_\times Q_\times , F_+ Q_+) -\pi/4 ,
\end{equation}
and where 
\begin{equation}
	Q\equiv \sqrt{ (F_+ Q_+)^2 + (F_\times Q_\times)^2}.
\end{equation}
If we replace $Q$ with its root mean square (see e.g. \S~7.7.2 in \cite{Maggiore:2007ulw})
\begin{equation}
	Q\rightarrow \sqrt{\langle Q^2 \rangle} = 4/5\,,
\end{equation} 
where the averaging symbol stands for 
\begin{equation}
	\langle \cdot \rangle =\int_{-1}^1 \frac{d\cos \iota}{2} \int_{-1}^1 \frac{d\cos \theta_S}{2}\int_0^{2\pi}\frac{d\phi_S}{2\pi}\int_0^{2\pi}\frac{d\psi_S}{2\pi},
\end{equation}
and rewrite the frequency dependence through $u=\nu^{1/5}x_f^{1/2}$, the result is the ppE waveform stated in Eqs.~\eqref{simplestppE}-\eqref{leadingTaylorF2} with 
\begin{align}
	\alpha &=\zeta \nu^{-2k/5}H\\ \label{alpha_ppE}
	&= \zeta \nu^{-2k/5} \big[ \frac{C}{2}(\delta_{k,p}+\delta_{k,q}-\delta_{p,q})+2\lambda \delta_{k,p} \big] ,\\ \label{a_ppE}
	a&=2k ,\\  
	\beta &= \zeta \frac{3}{128}\nu^{-2k/5}W\\ \label{beta_ppE}
	&=\zeta \nu^{-2k/5}\frac{15}{32}\frac{C}{(k-4)(2k-5)}  ,\\ 
	b&=2k-5 \label{b_ppE} .
\end{align}

%%%%%%%%%%%%%%%%%%%%%%%%%%%%%%%%%%%%%%%%%%%%%%%%%%%%%%%%%%%%%%%%%%%%%%%%%%%
\section{A New higher-harmonics ppE waveform model}
\label{sec:NewppE}

In this section, we extend the ppE waveform template in Eq.~\eqref{simplestppE} to incorporate harmonics beyond the dominant quadrupolar mode. We then propose a minimal model that can be used to asses how much the presence of new of amplitude corrections can impact parameters estimation. Despite its simplicity, the minimal Fourier-space waveform model is able to capture the class of modified theories that exhibit dipolar emission. Two examples, for which GW results to 1PN order have been computed, are shown: shift-symmetric-Gauss-Bonnet gravity and massless Scalar-Tensor theories.

%----------------------------------------------------------------------------
\subsection{A Higher-Mode ppE Model}
Our goal is to parameterize a Fourier-domain template that covers all the nine positive-frequency independent $\ell=2,3,4$ harmonics generated by an inspiraling non-spinning binary in quasi-circular orbit, for which the equatorial symmetry in Eq.~\eqref{h_equatorial_symmetry} holds. The main difficulty consists in narrowing down the number of ppE parameters that enables a mapping to different modified theories and that, at the same time, avoids overfitting the data.

First, consider the trivial generalization of Eq.~\eqref{h_GR_harmomics}
\begin{equation}\label{higher_harmonic_ppE_v1}
	\tilde{h}_{\ell m}(f) = \tilde{h}_{\ell m}^\GR(f) (1+ \alpha_{\ell m} u^{a_{\ell m}})\exp\big[{i\beta_{\ell m} u^{b_{\ell m}}}\big] ,
\end{equation}
with $u = (\pi \mathcal{M}f)^{1/3}$ and a maximal number of 36 constant parameters $\lbrace \alpha_{\ell m} ,a_{\ell m},  \beta_{\ell m},b_{\ell m} \rbrace $. Because the phase in the Fourier-domain is completely determined by inserting its time evolution $\dot{x}^{-1}(x)$ of Eq.~\eqref{oneoverxdot_ppE} within Eq.~\eqref{Fourier_phase}, we see that the $(\ell , m)$-dependence of the 18 parameters $\lbrace \beta_{\ell m},b_{\ell m} \rbrace$ simplifies to
\begin{equation}\label{beta_m_b}
	\beta_{\ell m} = \beta_m , \qquad b_{\ell m} = b ,
\end{equation}
where $b$ is the same as in Eq.~\eqref{b_ppE} and 
\begin{equation}
	\beta_{m} = \left(\frac{2}{m}\right)^{2(k-4)/3}\beta ,
\end{equation}
with $\beta$ given in Eq.~\eqref{beta_ppE}.

Is it sufficient to consider purely real amplitude corrections $\alpha_{\ell m}$?
Reference \cite{Islam:2019dmk} shows that the inclusion of a phase $\Theta_{\ell m} \equiv \text{arctan2}(\text{Im}\, \alpha_{\ell m},\text{Re}\, \alpha_{\ell m})$ is essentially uninformative when trying to place bounds on amplitude deviations. Therefore we restrict our analysis to the case $\alpha_{\ell m}\in \mathbb{R}$.  This choice implies that the ppE phase of Eq.~\eqref{higher_harmonic_ppE_v1} is entirely controlled by $\beta_m$ in Eq.~\eqref{beta_m_b} for all harmonics, or equivalently, by the knowledge of $\dot{x}^{-1}(x)$ to leading-PN order in the non-GR deformation.

As for the amplitude, we first identify the quadrupolar corrections $\lbrace \alpha_{22} , a_{22}\rbrace$ as the standard ppE amplitude parameters
\begin{equation}\label{map_ppE_quadrupole_amplitude}
    \alpha_{22} = \alpha, \qquad a_{22} = a ,
\end{equation}
with $\alpha$ and $a$ given by Eq.~\eqref{alpha_ppE} and Eq.~\eqref{a_ppE} respectively.
This can be seen by computing the Fourier-domain polarizations
\begin{align}
    \tilde{h}_+(f) &= \frac{1}{2}\big[ \tilde{\text{h}}(f) + \big(\tilde{\text{h}}(-f)\big)^* \big] ,\\
    \tilde{h}_\times(f) &= \frac{i}{2}\big[ \tilde{\text{h}}(f) - \big(\tilde{\text{h}}(-f)\big)^* \big],
\end{align}
at the detector's azimuthal position $\varphi =\pi /2$, where $\tilde{\text{h}}(f)$ is the Fourier transform of $\text{h}(t)$. 
By keeping the leading-PN harmonics $(\ell =2,m=\pm 2)$,  the only terms that survive the SPA are
\begin{align}
    \tilde{\text{h}}(f) &= \tilde{h}_{22}(f) \, Y_{-2}^{22}(\iota,\varphi) , \\
    \tilde{\text{h}}(-f) &= \big(\tilde{h}_{22}(f)\big)^* \, Y_{-2}^{2,-2}(\iota,\varphi) ,
\end{align}
where in the last equation we used the equatorial symmetry in Fourier space
\begin{equation}
    \tilde{h}_{\ell,-m}(-f) = (-1)^\ell \big( \tilde{h}_{\ell m}(f) \big)^* ,
\end{equation}
which follows from Eq.~\eqref{h_equatorial_symmetry}. The resulting response function, after discarding an immaterial overall phase and averaging over all angles, takes the standard ppE-form in Eq.~\eqref{simplestppE} from which Eq.~\eqref{map_ppE_quadrupole_amplitude} follows.

The crucial difference between the original ppE model in Eq.~\eqref{simplestppE} and the harmonic-enhanced version in Eq.~\eqref{higher_harmonic_ppE_v1} stems from the remaining amplitude corrections. 
As it is clear from the GR spherical-harmonic modes $h_{\ell m} (t)$ in Eq.~\eqref{h_ell_m_time_domain}, when $\ell >2$ the leading PN term in each harmonic is of order 0.5PN or higher, relative to the leading $h_{2 2}(t)$. This means that, unless the modified theory excites these harmonics through 0PN corrections---as it occurs when a scalar field sources the breathing mode---obtaining $\lbrace \alpha_{\ell m}, a_{\ell m}\rbrace_{\ell>2}$ requires the computation of non-GR $\mathcal{O}(\zeta)$ effects at next-to-leading PN accuracy. A consequence of this is that the analysis we used to handle the binary dynamics in Sec.\ref{sec:ppEnutshell} may not be sufficient to model such effects because there we discarded amplitude terms that directly contribute to higher harmonics.

Some exceptions to this argument, however, do exist. Reference~\cite{Shiralilou:2021mfl} recently calculated the waveform polarizations for quasi-circular binaries in scalar Gauss-Bonnet theory to 1PN order. In this theory, dipole radiation is excited in black hole binaries because individual black holes carry a monopolar scalar charged sourced by the Kretschmann scalar. Reference~\cite{Shiralilou:2021mfl} found that beyond leading-PN order terms can still be covered by the standard ppE formulation because one can model the 1PN terms as quadrupolar deformations, at the cost of introducing mismodelled terms that do not enter the relevant harmonics, as we will show in Sec.~\ref{sec:Mapping_known_theories}.

%----------------------------------------------------------------------------
\subsection{A minimal model}
\label{sec:Minimal_model}
In the absence of a straightforward link between the binary dynamics in modified theories and the ppE amplitude parameters $\lbrace \alpha_{\ell m}, a_{\ell m}\rbrace_{\ell>2}$, we can explore what happens in a model that mimics the amplitude choice we made at the end of Sec.~\ref{sssec:Time_domain_waveform_model}.
In this model, beyond-GR deformations couple to each mode through the amplitude and are insensitive to both polarization and PN-order. That is, the non-zero coefficients that define each amplitude in Eq.~\eqref{PN_amplitdue_decomposition} get stretched by the same frequency-dependent power
\begin{equation}
	a_{+,\times}^{(n,j)}\rightarrow a_{+,\times}^{(n,j)}\big(1+\zeta \,\Gamma^{(n)}\,x^{\gamma^{(n)}}\big) ,
\end{equation}
resulting in 
\begin{equation}\label{Minimal_model_amplitudes}
	A_{+,\times}^{(n)}(x)\rightarrow A_{+,\times}^{(n)}(x)\big(1+\zeta \,\Gamma^{(n)}\,x^{\gamma^{(n)}}\big) .
\end{equation}
This necessitates eight parameters $\lbrace\Gamma^{(n)},\gamma^{(n)}\rbrace_{n=1}^{4}$, on top of $\lbrace C, k \rbrace$, to describe all the $\ell = 2,3,4$ harmonics. The corresponding amplitude deformations in 
Eq.~\eqref{higher_harmonic_ppE_v1} do not depend on $\ell$ and read
\begin{align}\label{minimal_ppE}
	\alpha_{m} &= \zeta \left( \frac{2}{m}\right)^{\frac{2}{3}\sigma^{(m)}} \nu^{-\frac{2}{5}\sigma^{(m)}}\Big[ \frac{C}{2}\big(\delta_{\sigma^{(m)},\gamma^{(m)}}+ \\
	& +\delta_{\sigma^{(m)},k} -\delta_{\gamma^{(m)},k}\big)
	+\Gamma^{(m)} \delta_{\sigma^{(m)},\gamma^{(m)}} \Big] ,\\
	a_m & = 2 \sigma^{(m)} ,
\end{align}
where we have defined 
\begin{equation}
	\sigma^{(m)} \equiv \min \big(k, \gamma^{(m)} \big) .
\end{equation}
This model is analogous to the one explored in Sec.~IV of  \cite{Chatziioannou:2012rf} where each mode that is proportional to $\cos n \phi(t)$ receives an $n$-dependent modification. In that work, the ansatz in Eq.~\eqref{ansatz_ampl} is promoted to higher modes by conjecturing $A^{(n)}_{+,\times}\propto (r \dot{\phi})^n$ and subsequently a modified Kepler's law like Eq.~\eqref{kepler} is used.

Though the choice in Eq.~\eqref{Minimal_model_amplitudes} may not cover \textit{all} theories of gravity exactly, it remains a reasonable benchmark to test whether these amplitude corrections actually impact parameter estimation. Moreover, this minimal model is sufficient to cover all known 1PN waveform models in the modified gravity theories that have been worked out to date, as we show in the next subsection.

%----------------------------------------------------------------------------
\subsection{Mapping to known modified theories}
\label{sec:Mapping_known_theories}
We first consider shift-symmetric-Gauss-Bonnet (ssGB) gravity (see e.g. \cite{Sotiriou:2014pfa}), though the mapping to the ppE framework can be extended to the more general Einstein-scalar-Gauss-Bonnet (EsGB) theory \cite{Julie:2019sab,Shiralilou:2021mfl}. For ssGB gravity the dimensionless coupling is chosen to be
\begin{equation}
    \zeta_{\ssGB} = \frac{\alpha_{\ssGB}^2}{M^4} 
\end{equation}
where $\alpha_{\ssGB}$ is the fundamental coupling constant that enters in the action and $M$ is the total mass of the binary system (normalized by $16 \pi$, as in Eq. (II.4) of \cite{Julie:2019sab} or Eq.(3) of \cite{Witek:2018dmd}).

In ssGB theory, for two spinless compact objects in a circular orbit, the binding energy to leading-PN order, under the small coupling approximation, reads 
\cite{Yunes:2011we, Shiralilou:2021mfl,Julie:2019sab,Lyu:2022gdr,Bernard:2022noq}
\begin{equation}\label{Energy_ssGB}
    E = -\frac{\mu }{2} \hat{x} \left( 1 + \frac{4}{3}\frac{\zeta_\ssGB}{\nu^2}\hat{x} \right) ,
\end{equation}
where 
\begin{equation}
    \hat{x}\equiv (\bar{\alpha}_\ssGB M \omega)^{2/3} ,
\end{equation}
is the rescaled PN frequency parameter, in which the coupling-dependent constant 
\begin{equation}
    \bar{\alpha}_\ssGB\equiv 1+\frac{\zeta_\ssGB}{\nu^2} ,
\end{equation}
enters as a renormalization of the bare constant $G$ (set to $1$ here). Comparing Eq.\eqref{Energy_ssGB}
 to Eq.~\eqref{binding_E} we establish that 
 \begin{align}
     A &= \frac{4}{3\nu^2}, \\
     p &= 1 .
 \end{align}
 
In ssGB theory, the leading-PN-order flux in the small coupling limit is 
\begin{equation}\label{Flux_ssGB}
    \mathcal{F} = \frac{32 }{5}\nu^2 \hat{x}^5 \left( 1 + \zeta_\ssGB \frac{5}{96}\frac{\Delta^2}{\nu^4}\hat{x}^{-1} \right) ,
\end{equation}
with $\Delta = \text{sgn}(m_1-m_2) \sqrt{1-4\nu}$ and ($m_1 , m_2)$ the individual masses of the binary.
Comparing \eqref{Flux_ssGB} to Eq.~\eqref{flux}, allows the identification 
\begin{align}
     B &= \frac{5}{96}\frac{\Delta^2}{\nu^4}, \\
     q &= -1 .
 \end{align}
Then, using Eq.~\eqref{oneoverxdot_ppE_k_C}, we see that the parameters governing $\dot{x}^{-1}$ in Eq.~\eqref{oneoverxdot_ppE} are 
 \begin{align}
     C &= -\frac{5}{96}\frac{\Delta^2}{\nu^4}, \\
     k &= -1 .
 \end{align}
 
By taking the small coupling limit of the time-domain 1PN-order polarization waveforms in ssGB (see appendix E of \cite{Shiralilou:2021mfl}) we find the amplitudes $A^{(n),\ssGB}_{+,\times}$ of the theory to be
\begin{equation}\label{amplitude_ssGB}
    A^{(n),\ssGB}_{+,\times}(\hat{x}) =
    \begin{cases}
        A^{(n)}_{+,\times}(\hat{x}) &\text{if } n = 1,3,4 \\
        A^{(2)}_{+,\times}(\hat{x})(1 + \zeta_\ssGB \frac{4}{3\nu^2}\hat{x})  &\text{if } n = 2 .
    \end{cases}
\end{equation}
where $A^{(n)}_{+,\times}(\hat{x})$ is the GR amplitude in Eq.~\eqref{PN_amplitdue_decomposition}, truncated to 1PN order and evaluated at $x = \hat{x}$\footnote{In order to write $A^{(2),\ssGB}_{+,\times}(\hat{x})$ as in Eq.~\eqref{amplitude_ssGB}, we have approximated 
\begin{equation}
        a_{+,\times}^{(2,0)}(1+\zeta\, \Gamma^{(2)}\hat{x}) + a_{+,\times}^{(2,2)}\hat{x},
\approx \left( a_{+,\times}^{(2,0)} + a_{+,\times}^{(2,2)}\hat{x}\right)(1+\zeta \,\Gamma^{(2)}\hat{x})
\end{equation}
at the cost of introducing an uncontrolled remainder at ${\cal{O}}(\hat{x}^2)$, where $a_{+,\times}^{(n,j)}$ are the coefficients in App.~\ref{app:GR_waveform_amplitudes} that define the GR amplitudes $A^{(n)}_{+,\times}$ of Eq.~\eqref{PN_amplitdue_decomposition}. }. 
The truncation of the modification to the amplitude at 1 PN order will probably not affect the constraints one can place on the GR deformations, as found in the analogous phase study of~\cite{Perkins:2022fhr}; a verification of this confirmation is outside the scope of this paper. 
Note that, regardless of whether the waveforms used to extract Eq.~\eqref{amplitude_ssGB} are computed in the Einstein or the Jordan frame,
they coincide in the weak-field limit \cite{Shiralilou:2021mfl}, which we assumed here.

The amplitudes in Eq.~\eqref{amplitude_ssGB} are covered by the minimal model of Sec.~\ref{sec:Minimal_model} where $\Gamma^{(1)}=\Gamma^{(3)}=\Gamma^{(4)}=0$, 
\begin{equation}
    \Gamma^{(2)} = \frac{4}{3}\frac{1}{\nu^2}, \qquad \gamma^{(2)} = 1,
\end{equation}
and $\lbrace \gamma^{(m)} \rbrace_{m=1,3,4}$ can take any value. As this case exemplifies, whenever $k<0$ and the non-GR amplitude corrections in the time domain are of positive PN order, $\gamma^{(m)}>0$, the values of the coefficients ${\lbrace\Gamma^{(n)}}\rbrace_{n=1}^4$ do not enter $\alpha_m$ because the corrections to $\dot{x}^{-1}$ are dominant. The argument can be applied to theories whose time-domain amplitudes carry only positive PN-order correction, even if these are not described by Eq.~\eqref{Minimal_model_amplitudes}. 

In Massless Scalar-Tensor (ST) theories (see e.g. \cite{Bernard:2022noq, Sennett:2016klh}) the Fourier-domain GW modes $\tilde{h}_{\ell m}$ can still be parameterized by the minimal model $\lbrace \alpha_m , a_m\rbrace$ of Eq.~\eqref{minimal_ppE} despite the time-domain amplitude corrections being $\ell$-dependent. For ST theories, the dimensionless coupling is taken to be
\begin{equation}
    \zeta_{\ST} = \frac{1}{4 + 2\omega_0},
\end{equation}
where $\omega_0$ is the function coupling of the theory evaluated at the asymptotic value of the scalar field.
The PN expansion parameter is
\begin{equation}
    \tilde{x} = ( \bar{\alpha}_\ST M \omega)^{2/3}
\end{equation}
and the coupling-dependent rescaling $\bar{\alpha}_\ST$ of the gravitational constant is 
\begin{equation}
    \bar{\alpha}_\ST = 1- \zeta_\ST +\zeta_\ST (1-2s_1)(1-2s_2)
\end{equation}
with $s_1$ and $s_2$ the sensitivities of the compact objects.

In massless ST theories, and under the small coupling approximation, the structure of the binding energy (Eq. (5.5) of \cite{Bernard:2022noq}) and the flux (Eq. (5.3) of \cite{Bernard:2022noq}) for circular orbits is the same as in Eq.~\eqref{binding_E} and Eq.~\eqref{flux}, respectively. The coefficients are
\begin{align}
     A &= 16 (1-2s_1)(1-2s_2) , \\
     p &= 1 ,\\
     B &= \frac{5}{24}(s_1 - s_2)^2 , \\
     q &= -1 ,
 \end{align}
which lead to $C = - 5(s_1 - s_2)^2/24$ and $k = -1$ for the evolution of $\tilde{x}$. 

The time-domain waveform modes $h_{\ell m}(t)$ computed in the Jordan frame (see Sec.~VI of \cite{Sennett:2016klh}) exhibit only positive PN-order corrections, which are subdominant to the -1PN-order term in the flux. Thus, massless ST theories are still controlled by the minimal model of Sec.~\ref{sec:Minimal_model} in the small coupling regime, which is expected due to the known mapping to ssGB \cite{Julie:2019sab,Lyu:2022gdr}.

The waveform in \cite{Sennett:2016klh} shows also an overall 0PN-order factor $(1-\zeta_\ST)$, but this can be reabsorbed into a redefinition of the luminosity distance $D$, similarly to a redshift effect.
If the binary system contains only one BH ($s_2\neq s_1 =1/2$) then $\bar{\alpha}_\ST = 1 - \zeta_\ST$ and the binding energy receives a leading 2PN-order non-GR correction ($p=2$), which does not alter the values of $\lbrace C , k\rbrace$ above. For such systems, the overall 0PN-order factor coincides with $\bar{\alpha}_\ST$ and so it can be exactly reabsorbed into the rescaling of the bare constant $G$ (or equivalently the total mass $M$). 

%%%%%%%%%%%%%%%%%%%%%%%%%%%%%%%%%%%%%%%%%%%%%%%%%%%%%%%%%%%%%%%%%%%%%%%%%%%
\section{Conclusions}
\label{sec:Conclusions}

We have here summarized the features of a particular dominant-mode parameterized model in the frequency domain, namely the ppE waveform model, and then proposed a simple extension of it that covers subdominant modes. The latter can be particularly important for certain binary configurations \cite{Divyajyoti:2021uty}, and thus, the higher-harmonic extension of the ppE model may lead to more stringent constraints in the future.  The main difference between the new model and the basic ppE model is the presence of new amplitude corrections, which can be mapped to specific gravity theories as shown in the case of ssGB and ST theories.

The higher-harmonic ppE model, once implemented in a full waveform generator, could be  used to run a Bayesian analysis on known gravitational wave events from the LIGO and
Virgo gravitational wave catalogs. The output would determine the impact of the newly introduced amplitude corrections on constraints over theory-specific couplings, and their dependence on binary properties (both intrinsic, like masses and spins, and extrinsic, like distance and orientation). But how would one go about implementing this extended ppE model in a full waveform generator? 

An implementation is possible with phenomenological, or \textit{phenom}, models: analytical GW templates built to accurately describe the inspiral, merger, and ringdown of BH binaries in the frequency domain \cite{Ajith:2007kx, Ajith:2007qp, Ajith:2009bn, Santamaria:2010yb, Hannam:2013oca}. 
In the phenom construction, a waveform ansatz, containing a set of free phenomenological coefficients, is chosen for the inspiral-merger and ringdown phases. The structure of the frequency dependence is well motivated by PN theory and Quasi-Normal Mode (QNM) theory. Then, the free coefficients are fixed so that the waveform ansatz is both smooth at the inspiral-merger interface and it fits a chosen catalog of target waveforms, which cover the parameter space.

Many versions of the phenom models exist. For non-precessing sources, the dominant multipole model \texttt{IMRPhenomD} \cite{Husa:2015iqa, Khan:2015jqa} performs well as an inexpensive tool to generate accurate waveforms that can be employed in GW searches. Its extension, \texttt{IMRPhenomHM} \cite{ London:2017bcn}, incorporates beyond-quadrupole radiative moments to mitigate the bias on the properties of edge-on and asymmetric binaries inferred through parameter estimation.\footnote{See \cite{Mehta:2017jpq} for a model analogous to \texttt{IMRPhenomHM}.}

An implementation of our extended ppE model on \texttt{IMRPhenomHM} is straightforward because of the latter's modularity. That is, the GR inspiral portion of \texttt{IMRPhenomHM} could be straightforwardly replaced with the ppE counterpart in Sec.~\ref{sec:Minimal_model}. This would require adding ppE parameters to the parameter space, and then, given a specific theory and its mapping to the ppE parameters, re-calibrating the phenomenological coefficients against hybrid target waveforms that are produced by that modified theory. However, numerical waverform catalogs in modified gravity are currently scarce at best, so in practice the tuning of phenomenological coefficient has to be done using the GR catalogs. Because our extended ppE model is theory-agnostic, the latter choice is expected to at most produce conservative bounds on known modified theories.

Finally, we note that by requiring that the amplitude and phase match smoothly at the inspiral-merger frequency, the merger segment of the model will inherit ppE corrections accumulated during the many cycles of the inspiral portion. Thus, the merger is deformed even without including a complete QNM analysis of each theory, which would shift the QNM frequencies (see e.g.~\cite{Wagle:2021tam}). Adding the latter is expected to strengthen constraints on modified gravity, and it could be pursued as a follow up analysis to this paper.

The preliminary model in Sec.~\ref{sec:Minimal_model} can be improved in different aspects, suggesting future projects that might be worth exploring. First, one could include higher-PN order non-GR corrections in each $\tilde{h}_{\ell m}(f)$ and estimate how these affect the marginalized posteriors in parameter estimation. If not available in the literature, these corrections can still be inserted artificially by proposing a dependence on the masses (e.g.~casting them as a quadratic polynomial in the mass ratio) and adding more phenomenological coefficients to control their magnitude. Second, one could employ genuine QNM results for each theory to refine the ringdown regime ansatz. This would offset the damping frequencies and the values of the frequencies where the smooth matching is done. Lastly, if the initial analysis with the higher-harmonic model proves promising, one could invest in generating hybrid waveform catalogues for modified gravity theories whose PN form is known. Using these more appropriate catalogues to fix the phenomenological coefficients would minimize the unfaithfulness and possibly yield stronger reliable bounds on the coupling constant of the theories we test.

%%%%%%%%%%%%%%%%%%%%%%%%%%%%%%%%%%%%%%%%%%%%%%%%%%%%%%%%%%%%%%%%
\section*{Acknowledgments}

We thank Scott Perkins for useful
comments. N.Y. acknowledges financial support through NASA ATP Grants No. 17-ATP17-0225, No. NNX16AB98G and No.80NSSC17M0041.

%%%%%%%%%%%%%%%%%%%%%%%%%%%%%%%%%%%%%%%%%%%%%%%%%%%%%%%%%%%%%%%%
\appendix

%----------------------------------------------------------------
\section{GR waveform amplitudes}
\label{app:GR_waveform_amplitudes}
Here we list the coefficients defining the GR amplitudes $\lbrace A^{(n)}_{+,\times} (x)\rbrace_{n=1}^5$ to 1.5PN ($j\leq 3$), used in Sec.~\ref{sec:Harmonic_decomposition}.

Denoting $c_\iota\equiv \cos\iota$, for the plus polarization we have
\begin{subequations}
	\begin{equation}
		a^{(1,1)}_+ = \frac{1}{8}   \sqrt{1-c_\iota ^2} \left(c_\iota ^2+5\right) \Delta ,
	\end{equation}
	\begin{align}
			a^{(1,3)}_+ =&\frac{1}{192}\sqrt{1-c_\iota ^2}\big[(1-2 \nu ) c_\iota ^4+ \\ \nonumber
			&-(24 \nu +60) c_\iota ^2+98 \nu -57 \big] \Delta ,
	\end{align}
	\begin{align}
		a^{(2,0)}_+ &= c_\iota^2 +1  ,
	\end{align}
	\begin{align}
		a^{(2,2)}_+ &= \frac{1}{6} \big[2 (1-3 \nu ) c_\iota ^4-(11 \nu +9) c_\iota ^2+19 (\nu -1)\big] ,
	\end{align}
	\begin{align}
		a^{(2,3)}_+ &= 2 \pi  \left(c_\iota ^2+1\right) ,
	\end{align}
	\begin{align}
	a^{(3,1)}_+ &= -\frac{9}{8}   \sqrt{1-c_\iota ^2} \left(c_\iota ^2+1\right) \Delta,
	\end{align}
	\begin{align}
	a^{(3,3)}_+ &=\frac{9}{128}  \sqrt{1-c_\iota ^2} \big[ (18 \nu -9) c_\iota ^4+ \\ \nonumber
	&+(16 \nu +40) c_\iota ^2-50 \nu +73\big] \Delta ,
	\end{align}
	\begin{align}
	a^{(4,2)}_+ &= \frac{4}{3} (3 \nu -1) (c_\iota^4 -1) ,
	\end{align}
	\begin{align}
	a^{(5,3)}_+ &= \frac{625}{384}  (1-2 \nu) \sqrt{1-c_\iota ^2} (c_\iota^4 -1)  \Delta ,
	\end{align}
\end{subequations}
whereas for the cross polarization
\begin{subequations}
	\begin{equation}
		a^{(1,1)}_\times = \frac{3}{4}   c_\iota  \sqrt{1-c_\iota ^2} \Delta,
	\end{equation}
	\begin{align}
		a^{(1,3)}_\times =&\frac{1}{96}   c_\iota  \sqrt{1-c_\iota ^2} \big[5 (1-2 \nu ) c_\iota ^2+46 \nu -63\big] \Delta ,
	\end{align}
	\begin{align}
		a^{(2,0)}_+ &= 2c_\iota ,
	\end{align}
	\begin{align}
		a^{(2,2)}_\times &= \frac{1}{3} c_\iota  \big[4 (1-3 \nu ) c_\iota ^2+13 \nu -17\big] ,
	\end{align}
	\begin{align}
		a^{(2,3)}_\times &= 4 \pi  c_\iota ,
	\end{align}
	\begin{align}
		a^{(3,1)}_\times &= -\frac{9}{4}  c_\iota  \sqrt{1-c_\iota ^2} \Delta ,
	\end{align}
	\begin{align}
		a^{(3,3)}_\times &=\frac{9}{64}  c_\iota  \sqrt{1-c_\iota ^2} \big[15 (2 \nu -1) c_\iota ^2-38 \nu +67\big]\Delta  ,
	\end{align}
	\begin{align}
		a^{(4,2)}_\times &= \frac{8}{3} (3 \nu -1) c_\iota  \left(c_\iota ^2-1\right) ,
	\end{align}
	\begin{align}
		a^{(5,3)}_\times &= \frac{625}{192}   (2 \nu -1) c_\iota  \left(1-c_\iota ^2\right)^{3/2} \Delta ,
	\end{align}
\end{subequations}
and all the other coefficients are zero.
\\
%----------------------------------------------------------------
\section{Spin-weighted $s=2$ spherical harmonics}
\label{app:Spin2_Spherical_Harmonics}
Here we summarize the definition and the properties of the spin-2 spherical harmonics, used in the harmonic decomposition. Following the conventions in \cite{Arun:2008kb}, we have
\begin{equation}\label{Spin2_Spherical_Harmonics_Y}
    Y_{-2}^{\ell m}(\iota , \varphi ) = \sqrt{\frac{2\ell +1}{4\pi}}\, d^{\ell m}_2(\cos \iota)\, e^{i m \varphi} ,
\end{equation}
where
\begin{align} \label{Spin2_Spherical_Harmonics_d}
    d^{\ell m}_2(\cos \iota) &= \sum_{k\in I_{\ell m}} c_{k,\ell m}\, (1+\cos\iota)^{\ell +m/2 -k-1}\times \\ \nonumber 
    &\times (1-\cos\iota)^{-m/2 +k+1}  , 
\end{align}
and the indices $(\ell ,m)$ take integer values $\ell\geq 2 $, $|m|\leq \ell$.
The coefficients  $c_{k,\ell m}$ are defined as 
\begin{equation}\label{Spin2_Spherical_Harmonics_d_c}
    c_{k,\ell m}=\frac{(-1)^k}{k!}\frac{2^{-\ell}\sqrt{(\ell +m)!(\ell -m)!(\ell+2)!(\ell -2)!}}{ (k-m+2)!(\ell +m-k)!(\ell -k-2)!} ,
\end{equation}
and the $k$-sum is over the subset of integers specified by
\begin{equation}\label{Spin2_Spherical_Harmonics_I}
    I_{\ell m} = \lbrace k \in \mathbb{Z}\, | \max(0, m-2) \leq k \leq \min(\ell + m, \ell -2) \rbrace .
\end{equation}
From these definitions follow the orthonormality relation
\begin{equation}\label{Spin2_Spherical_Harmonics_Orthon}
    \int_{-1}^1 d\cos\iota \int_0^{2\pi} d\varphi \,\, Y_{-2}^{\ell m}(\iota , \varphi ) \left( Y_{-2}^{\ell' m'}(\iota , \varphi ) \right)^* = \delta_{\ell \ell'}\delta_{m m'} ,
\end{equation}
the completeness relation
\begin{equation}\label{Spin2_Spherical_Harmonics_Complete}
    \sum_{\ell =2}^{\infty}\sum_{m=-\ell}^{\ell} Y_{-2}^{\ell m}(\iota , \varphi ) \left( Y_{-2}^{\ell m}(\iota' , \varphi') \right)^* = \delta(\cos\iota -\cos\iota')\delta(\varphi-\varphi'),
\end{equation}
and the symmetry
\begin{equation}
    d_2^{\ell , -m}(\cos \iota) = (-1)^\ell d_2^{\ell m}(-\cos\iota).
\end{equation}
The latter enables the equatorial symmetry in Eq.~\eqref{h_equatorial_symmetry} when the GW amplitudes $A^{(n)}_{+}$ and $A^{(n)}_{\times}$ are respectively even and odd in $\cos\iota$. 

%%%%%%%%%%%%%%%%%%%%%%%%%%%%%%%%%%%%%%%%%%%%%%%%%%%%%%%%%%%%%

\bibliography{references}

\end{document}